\documentclass[11pt,a4paper]{article}
\usepackage[english]{babel}
\usepackage{amssymb,amsmath}
\usepackage[latin1]{inputenc}
\usepackage[pdftex,bookmarks]{hyperref}
\usepackage{pdflscape}
\usepackage{anysize}
\usepackage[all]{xy}
\marginsize{3cm}{3cm}{3cm}{3cm}
\title{\huge{Recovering from Derivatives Funding} \\
\large{A consistent approach to DVA, FVA and Hedging}}

\author{C. Johan Gunnesson\thanks{cje.gunnesson@bbva.com}\qquad Alberto Fern\'andez Mu\~noz de Morales\thanks{a.fernandez.munozmor@bbva.com}\\ \\BBVA - Risk Methodologies\thanks{The opinions of this article are those of the authors and do not reflect in any way the views or business of their employer.}}

\begin{document}

\maketitle

\begin{abstract}

The inclusion of DVA in the fair-value of derivative transactions has now become standard accounting practice in most parts of the world. Furthermore, some sophisticated banks are including an FVA (Funding Valuation Adjustment), but since DVA can be interpreted as a funding benefit the oft-debated issue regarding a possible double-counting of funding benefits arises, with little consensus as to its resolution. One possibility is to price the derivative by replication, by constructing a portfolio that completely hedges all risks present in the instrument, guaranteeing a consistent inclusion of costs and benefits. However, as has recently been noted, DVA is (at least partially) unhedgeable, having no exact market hedge. Furthermore, current frameworks shed little light on the controversial question, raised by Hull (2012), of whether the effect a derivative has on the riskiness of an institution's debt should be taken into account when calculating FVA.

In this paper we propose a solution to these two problems by identifying an instrument, a fictitious CDS written on the hedging counterparty which, although not available in the market for active hedging, is implicitly contained in any given derivatives transaction. This allows us to show that the hedger's unhedged jump-to-default risk has, despite not being actively managed, a well-defined value associated to a funding benefit. Carrying out the replication including such a CDS, we obtain a price for the derivative consisting of its collateralized equivalent, a CVA contingent on the survival of the hedger, a contingent DVA, and an FVA, coupled to the price via the hedger's short-term bond-CDS basis. 

The resulting funding cost is non-zero, but substantially smaller than what is obtained in alternative approaches due to the effect the derivative has on the recovery of the hedger's liabilities. Also, price agreement is possible for two sophisticated counterparties entering a deal if their bond-CDS bases obey a certain relationship, similar to what was first obtained by Morini and Prampolini (2010).

\end{abstract}

\newpage

\setlength{\baselineskip}{1\baselineskip}

\section{Introduction}

Since its inception, the inclusion of Debit Valuation Adjustment in the pricing of derivatives has been controversial (see, for example, \cite{Gregory2009,Brigo2009}). Apart from issues related to monetization, and a skewed incentive structure, the debate has often focused on the impossibility of completely hedging DVA (\cite{Castagna2012}), due to hedgers not being able to freely trade their own Credit Default Swaps. Furthermore, much attention has recently been brought to Funding Valuation Adjustments, which include the hedger's funding costs in the pricing, with the consequence that care must be taken not to double count the part of FVA corresponding to a funding benefit with those implicit in DVA\footnote{These issues have been explored in, for example, \cite{PallavaciniBrigo2011} or \cite{PallavaciniBrigo2012}. For an introduction to these topics, see for instance, \cite{Brigo2013} or \cite{Gregory2012}}. Despite early approaches, such as \cite{MoriniPrampolini2010, PallavaciniBrigo2011}, for reconciling the two quantities, the issue has not been settled, and currently a wide array of possibilities exist. There is a debate around whether FVA should be included in the derivative's fair value from an accounting point of view, but from an economic perspective it would at first glance seem to make sense, in particular when considering hedging costs to be charged to the counterparty.

One way to guarantee a consistent inclusion of all risks and benefits is to price derivatives through replication, in which a portfolio of instruments, whose price is known, is constructed reproducing the pay-off of the original deal. With the same pay-off, the price of such a portfolio must equal that of the deal. As summarized in \cite{KenyonStamm2012}, such an analysis, including CVA, DVA and FVA, was first performed by \cite{BurgardKjaer2011} (see also \cite{BurgardKjaer2012}), where own credit risk, including jump-to-default risk, was replicated through the dynamic trading of a pair of bonds.

It has, however, been suggested that since DVA cannot be hedged (and it is frequently not even desirable to do so), the replicating portfolio should not hedge own jump-to-default risk, leading to proposals such as \cite{Garcia2013}, for pricing an uncollateralized transaction. By excluding this risk-factor from the replication, the price obtained does not depend on the hedger's jump-to-default and consequently DVA is absent from the pricing formula.

In this paper we will argue that, despite not being completely hedgeable, DVA has a well-defined value for the hedger, in terms of a funding benefit, prompting a modification of the replication argument. We will show that the value of the hedger's jump-to-default component can be modelled in terms of a fictitious CDS written on herself, which is implicitly sold to the hedger by her counterparty, and that can therefore be included in the replicating portfolio. Furthermore, we explain why the hedger should be willing to pay precisely the corresponding market spread for such a CDS. As a result, we obtain a pricing equation containing a CVA contingent on the survival of the hedger, a contingent DVA, and an FVA, coupled to the price via the hedger's bond-CDS basis.

This result sheds some light on the FVA = 0 debate (started in \cite{Hull}) since it is based on the interaction between a derivative's riskiness, from the view-point of credit worthiness, and the recovery of the rest of the hedgers' liabilities. The funding valuation adjustment does not vanish, however, and is instead governed by the bond-CDS basis, reflecting the fact that financing spreads contain considerations, such as liquidity, beyond mere quantification of credit risk. This possibility was first stated in \cite{Hull2012b}, and then further discussed in \cite{Hull2013}, under simpler assumptions than the ones that we consider.

Our analysis is based on the techniques used in \cite{BurgardKjaer2011} and \cite{BurgardKjaer2012}, incorporating stochastic credit spreads following \cite{Garcia2013},  where derivations for the equations governing the prices of credit instruments can be found. Also, we adopt a setup similar to that described in \cite{Garcia2013}, with the properties:

\begin{itemize}
\item The parties involved in the derivatives transaction are separated into a Hedger (H) and an Investor (I). The replication will be analyzed from the viewpoint of the hedger, while the sign of its value is as seen by the investor. The same valuation would not be obtained if it were the investor performing the replication, and the pricing will therefore be agent-specific.
\item The hedgers own credit spreads are allowed to be stochastic, in contrast with the deterministic spreads assumed in \cite{BurgardKjaer2011}.
\item The portfolio should be self-financing, with no additional funding obtained. This is based on the observation that if we hedge our own credit spread risk by buying back issued bonds, and we need to finance such a buy-back, the funding will imply issuing new bonds, subject to their own credit events. So the net effect is that it is not possible to hedge own jump-to-default by financed bond buy-backs. The solution of \cite{Garcia2013} is to hedge own spread risk, but not jump-to-default, by trading in two bonds $B(t,\,t+dt)$ and $B(t,\,T)$, of different maturities, satisfying the \textit{funding constraint\footnote{This is similar to equation (7) of \cite{BurgardKjaer2012}, with the difference that in our case, only the purchase of the hedger's bonds must be financed as the rest of the replicating portfolio will be fully collateralized.}}:

\begin{equation}
V_t = \Omega_t B(t,\,t+dt) + \omega_t B(t,\, T) \ ,
\label{eq:SFConstraintLM}
\end{equation}

where $V_t$ is the value of the derivative, and $\Omega_t$ and $\omega_t$ are the quantities held in the replicating portfolio of the short-termed and long-termed bond, respectively.
\item We assume a riskless close-out in the event that either hedger or investor defaults, implying that claims put forth by either the surviving party or the liquidators of the defaulted party will be made upon a MtM based on a perfectly collateralized transaction. No further defaults are thus considered when determining the close-out amounts. The analysis could be extended by incorporating more general boundary conditions (see \cite{Kjaer2011}, \cite{BurgardKjaer2012}), but for uncollateralized transactions, such as the one that we are concerned with here, a riskless closeout seems more appropriate. The reason is that the investor is likely to be less sophisticated than in the collateralized case, and possibly lacking the capability to incorporate CVA and DVA in close-out calculations, making it more probable that the simpler, riskless close-out is adopted.

\item Risk factors are, for simplicity, taken to be driven by single factor models, such as a single credit spread factor, and Interest rates are deterministic. Both of these assumptions can be relaxed without changing the final results.

\end{itemize}

\section{The Replicating Portfolio with the Investor entering a mandatory CDS}

In this section we will describe the replicating portfolio, explaining its composition in terms of a set of actively managed instruments, together with an unhedged component, which nevertheless has a well-defined value.

\subsection{Pricing one's own default}
Let us start by considering a short-term CDS, of notional $M$, written on the hedger, denoted $CDS^H(t,\,t+dt)$. This will lead us to argue that an unhedged jump-to-default risk will have a well-defined value for the hedger. Let us begin by studying the case where the CDS spread is zero, in which case the value of the CDS will be paid upfront. In the event that the hedger defaults between $t$ and $t+dt$, such a product will pay $M \cdot (1-R_H)$, where $R_H$ is the recovery fraction of the hedger, and 0 otherwise.

If the hedger were to price the purchase of this CDS, written on herself, she could not do so by replication using only tradable instruments, since it is not possible to perfectly hedge one's own jump-to-default. Does this mean that the value, from the viewpoint of the hedger, of such an instrument is zero? If it is, this implies that if the hedger were given the option to buy an upfront (zero-spread) short-term CDS written on herself, she should offer at most zero to do so.

But the purchased CDS will have positive value to the hedger. The reason is that in the event of default, the payout from the CDS will be distributed among the creditors of the hedger, raising the recovery amount they are able to secure. This will, in turn, lead to a reduction in the hedger's funding spread, since bondholders with the higher recovery in mind will accept lower interest payments\footnote{In the same way that debt of higher seniority will pay lower rates.}. So, a CDS on oneself, with zero spread, has a positive value due to the funding benefit it generates. In other words, if, instead of an upfront, the CDS is purchased through the payment of a spread, the hedger should be prepared to offer a non-zero spread.

In fact, we will now argue that the hedger should offer precisely her short-term CDS spread $\pi_t^H$, i.e. the spread that other market participants are offering for short-term protection on the hedger.

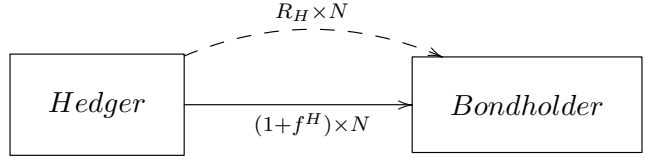
\begin{figure}[h]
\xymatrix@=3cm{
&*+<1cm>[F]{Hedger} \ar[r]_{(1+f^{H})\times N} \ar@/^1cm/@{-->}[r]^{R_{H}\times N} & *+<1cm>[F]{Bond holder}\\
}
\textbf{\caption{\small{Stylized cashflows on maturity of a bond issued by the hedger. The flow marked by the continuous line takes place if the hedger does not default, while the dashed line corresponds to the payment in the event of default.}}}
\label{fig:FlowsWithout}
\end{figure}

\begin{figure}[h]
\xymatrix@=3cm{
&*+<1cm>[F]{Hedger} \ar[r]_{(1+f^{H*})\times N} \ar@/^1cm/@{-->}[r]^{R_{H}\times N} \ar@/_1cm/@{-->}[r]_{M\cdot (1-R_{H})} & *+<1cm>[F]{Bond holder}\ar@/^1cm/@{-->}[d]^{M\cdot(1-R_{H})}\\
& & *+<1cm>[F]{Market} \ar@/^1cm/[u]^{M\cdot\pi^{H}}\\
}
\textbf{\caption{\small{Stylized cashflows at maturity of a bond issued by the hedger, when the hedger has a CDS of notional $M$ written on herself in her portfolio. In the event of the hedger's default the bondholder will, in the capacity of the hedger's sole creditor, receive the payoff from this CDS, which she may choose to hedge by selling a (different) CDS written on the hedger in the market. For simplicity, $R_H$ refers to the recovery corresponding to the previous case (without the extra cash-flow), which would increase the recovery.}}}
\label{fig:FlowsWith}
\end{figure}
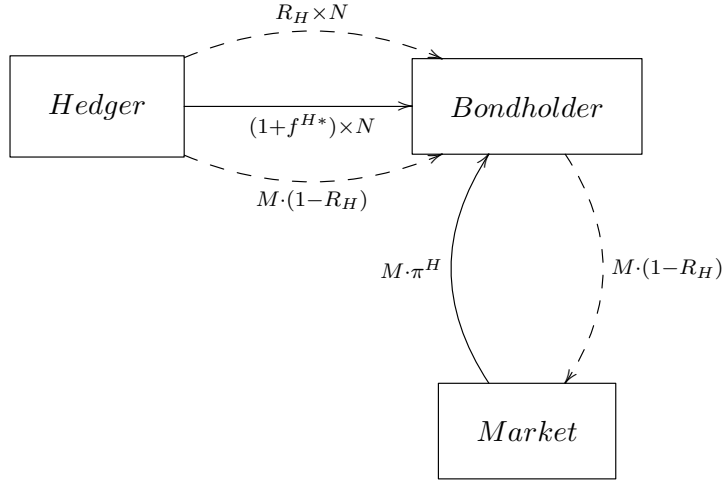

Let us consider two scenarios, both in which the hedger has issued, at an earlier stage, bonds of total notional $N$, comprising the totality of the hedger's liabilities. In the first scenario, depicted in Figure 1, the hedger simply pays, at maturity of the bond, an interest of $f^H\times N$ and returns the notional amount. In the case of the hedger's default, the bondholders receive only a recovery fraction $R_H$ of the notional.

Let us now imagine a different situation, shown in Figure 2, in which the hedger has previously bought a CDS on herself of notional $M$ that will payout $M\cdot (1-R_H)$ upon her default. This amount would then be distributed among her creditors, which, in this case, is comprised entirely of the bondholders. We denote the interest paid by the hedger in this second scenario $f^{H*}\times N$. Now,

\begin{itemize}
\item For simplicity, let us assume that the outstanding net liabilities corresponding to the derivatives portion of the hedger's balance-sheet are negligible compared to issued bonds, so that, in the event of default of the hedger, the bondholders will receive the complete CDS payout of $M\cdot (1-R_H)$. Furthermore, if $M<<N$, the recovery determining the CDS payout will not be affected by the payout itself. 
\item This will increase the hedger's recovery rate marginally, but the bondholders can also choose to hedge this default-contingent cashflow by selling protection, of total notional $M$, on the hedger in the standard market. They will then receive the market CDS spread times notional $M\cdot \pi^H$.
\item Due to this additional revenue, the bondholders will accept a reduction in the interest paid by the hedger by precisely $M\cdot \pi^H$, i.e., $f^{H*}\times N=f^H\times N - M\cdot \pi^{H}$. Summed over all bondholders, on entering the CDS position the hedger will reduce his interest payments by $M\cdot \pi^H$, while the situation for the bondholders will remain unchanged\footnote{Notice that the funding rate itself will not decrease by $\pi^H$, but that the aggregate interest payments decrease by $M$ times this spread.}.
\item In order to receive this funding benefit, the hedger would therefore be prepared to pay precisely $M\cdot \pi^H$. 
\end{itemize}

Note that it is not necessary for an actual default to take place in order to monetize this funding benefit since it is immediately realized\footnote{This funding benefit is similar to the one experienced by an ordinary mortgage taker, who may be given the option by financial entities to buy life insurance, which cancels the mortgage in the event of the taker's death. Obviously, the mortgage taker will not be exposed to the change in value of his portfolio resulting from his death, but the fact that he owns this insurance clearly eases his funding costs (not considering the insurance premium), either through a reduction of the interest paid or simply by granting him the access to the mortgage.}. Our main point is that if a financial transaction (such as a derivatives contract) contains a default-contingent cashflow (e.g. the amount of the PV that the hedger ends up not paying), this cashflow will provide a funding benefit in life. This benefit should therefore be included when pricing derivatives.

This argument is directly applicable when the derivatives component of the hedger's liabilities is small, since then the default-contingent cash flow will be distributed among the bondholders and other lenders, but when applied to the jump-to-default component of a derivatives transaction it should be valid in general. The reason is that, by definition, $(1-R_H)$ is the proportion of the derivatives' negative PV that is not paid out to the derivatives counterparties, and when compared against a scenario excluding such cash-flows, is thus distributed entirely among non-derivative creditors.

\subsection{The replicating portfolio}

We now turn to the replicating portfolio. The hedger will construct it in the same way as in \cite{Garcia2013}. It will thus consist of a collateralized derivative $H_t$, used for hedging market risk, short- and long-term CDSs\footnote{The short-term (infinitesimal) CDSs are a theoretical construct (see \cite{Garcia2013} for more details) used here for simplicity, but in practice, the same hedging can be carried out by trading in two contracts of finite, but different, maturity} written on the Investor, $CDS^I(t,\,t+dt)$ and $CDS^I(t,\,T)$, for eliminating investor credit-spread and jump-to-default risk, issued bonds $B(t,\,t+dt)$, $B(t,\, T)$, satisfying the funding constraint (\ref{eq:SFConstraintLM}) and hedging own credit spread risk, as well as cash in collateral accounts. The collateral account will be described using a unit-of-account $C_t$, of constant value 1, which generates an annualized interest of $c_t$, given by precisely the rate paid on collateral.

We now turn to the hedger's jump-to-default component. If the hedger were able to trade freely her own CDS it would constitute a natural hedge. This trading is not possible, but we can still model the value of the unhedged risk by adding the infinitesimal CDS, $CDS^H(t,\,t+dt)$ to the replicating portfolio. The reason is that in the case of default of the hedger between $t$ and $t+dt$, the investor will take a loss corresponding to precisely the pay-off of such a CDS. Equivalently, we can think of the investor as entering a derivative which contains two-components: one which is replicated by a portfolio, actively managed by the hedger, and one which is a CDS sold by the investor to the hedger. As seen above, the hedger will pay a spread $\pi_t^H$ for such a CDS, even if it were valued in isolation, and we can therefore price the unhedged jump-to-default component accordingly.

In other words, the deal entered by the hedger and investor contains, implicitly, a CDS written on the hedger, which the investor is obliged to sell to the hedger. By enforcing the replication equation and the cancelling of all non-deterministic terms, we are simply making use of the fact that this implicit CDS component will automatically adjust so that the jump-to-default component will be taken into account dynamically. Furthermore, by including the spread $\pi_t^H$ as the drift of $CDS^H(t,\,t+dt)$, the investor is properly compensated for the CDS that he has sold to the hedger, while the hedger is paying a rate corresponding to the received funding benefit. If the hedger were to include a different spread in the replication setup, the obtained price would not reflect her true replication costs (and benefits).

In sum, the replicating portfolio that we will use is

\begin{equation}
\begin{array}{ll}
V_t = & \alpha_t H_t + \beta _t C_t + \xi_t CDS^I(t,\, T) + \epsilon_t CDS^I(t,\,t+dt) + \Omega_t B(t,\,t+dt) + \omega_t B(t,\, T)\\
& + \eta_t CDS^H(t,\,t+dt)\\
\end{array}
\label{eq:ReplicatingPortfolio}
\end{equation}

\section{Performing the Replication}

In order to perform the replication, for simplicity we will assume that the evolution of the relevant market variables under the real measure $\mathbb{P}$ is described as

\begin{equation}
\left\lbrace
\begin{array}{l}
dS_t=\mu_{t}^{S} S_{t}dt+\sigma_{t}^{S}dW_{t}^{S,\mathbb{P}}\\
d\pi_t^{I}=\mu_t^{I}dt+\sigma_{t}^{I} dW_{t}^{I,\mathbb{P}}\\
d\pi_t^{H}=\mu_t^{H}dt+\sigma_{t}^{H} dW_{t}^{H,\mathbb{P}}\\
\end{array}
\right.
\label{eq:PDynamics}
\end{equation}
where $S_t$ represents the price of the derivative's underlying asset at time $t$, while $\pi_t^I$ and $\pi_t^H$ are the short term CDS spread of the investor and hedger, respectively. These spreads are defined so that $CDS^{k}(t,t+dt)=0, k\in\{I,H\}$. $\mu_{t}^{S}$, $\mu_{t}^{I}$ and $\mu_{t}^{H}$ are the real world drifts of these processes, while $\sigma_{t}^{S}(t,S_{t})$, $\sigma_{t}^{I}(t,\pi_{t}^{I})$, $\sigma_{t}^{H}(t,\pi_{t}^{H})$ are their volatilities. Interest rates will be taken to be deterministic, although stochastic rates would not affect the final results.

The three processes will be correlated with time dependent correlations:
\begin{equation}
\begin{array}{ccc}
\rho_{t}^{S,I}dt=dW_{t}^{S,\mathbb{P}} dW_{t}^{I,\mathbb{P}}&\rho_{t}^{H,I}dt=dW_{t}^{H,\mathbb{P}} dW_{t}^{I,\mathbb{P}}&\rho_{t}^{S,H}dt=dW_{t}^{S,\mathbb{P}} dW_{t}^{H,\mathbb{P}}\\
\end{array}
\end{equation}

Two additional sources of uncertainty are described by the default indicator processes $N_{t}^{I,\mathbb{P}}=1_{\{\tau^{I}\leq t\}}$ and $N_{t}^{H,\mathbb{P}}=1_{\{\tau^{H}\leq t\}}$, with real world intensities $\lambda_{t}^{I,\mathbb{P}}$ and $\lambda_{t}^{H,\mathbb{P}}$, with $\tau^{I}$ and $\tau^H$ being the default times of the investor and the hedger, respectively.

Let us call $V_t$ the NPV of the derivative from the investor's perspective. We will consider that both the hedger and the investor are defaultable, so the uncollateralized derivative will experience variations according to the following risk factors:
\begin{itemize}
\item Market risk due to changes in $S_t$
\item Investor's spread risk due to changes in $\pi_t^{I}$
\item Investor's default event
\item Hedger's spread risk due to changes in $\pi_t^{H}$
\item Hedger's default event
\end{itemize}

As we have shown in the previous section, the last component, despite being an unhedged jump-to-default risk, has a well-defined value, and a complete replicating portfolio will consist of two parts, one which can be traded by the hedger herself, as well as a short-term CDS written on the hedger, and sold by the investor.

As mentioned above, the collateral account is described using the unit-of-account $C_t$, which generates an interest of $c_t dt$ between $t$ and $t+dt$. The coefficient $\Omega_t$ is fixed by the funding constraint \eqref{eq:SFConstraintLM}, and $\beta_t$ by the amount of collateral held in collateral accounts (equal to $-\alpha_t H_t - \xi_t CDS^I(t,\, T)$, since the infinitesimal CDS are worth zero). The rest of the coefficients can be freely chosen by the hedger so as to carry out the replication strategy.

We will now proceed in the standard way, equating the differential of \eqref{eq:ReplicatingPortfolio}, assuming a self-financing strategy, with the expression obtained by expanding $dV_t$ using It\^o's Lemma, and choosing the available coefficients so that the stochastic terms cancel. The remaining, deterministic terms then imply a differential equation for $V_t$. Since this procedure is fairly standard, for the sake of brevity, we will omit some steps, spelling out explicitly certain terms.

Conditional on both the investor and the hedger being alive at time $t$, the change in $V_t$ under \emph{every} path will be given by (applying It\^o's Lemma for jump diffusion processes)

\begin{equation}
\begin{array}{ll}
dV_{t}=&\mathcal{L}_{SIH}V_{t}dt+\frac{\partial V_t}{\partial S_t}S_{t}\sigma_{t}^{S}dW_{t}^{S,\mathbb{P}}+\frac{\partial V_t}{\partial \pi_{t}^{I}} \sigma_{t}^{I}dW_{t}^{I,\mathbb{P}}+\frac{\partial V_t}{\partial \pi_{t}^{H}} \sigma_{t}^{H}dW_{t}^{H,\mathbb{P}}\\
&+\Delta V_{t}^{I}dN_{t}^{I,\mathbb{P}}+\Delta V_{t}^{H}dN_{t}^{H,\mathbb{P}},\\
\end{array}
\label{eq:dVFromIto}
\end{equation}
where $\mathcal{L}_{SIH}V_{t}$ groups together deterministic terms.

On the other hand, by assuming a self-replicating trading strategy, taking the differential of \eqref{eq:ReplicatingPortfolio} gives\footnote{Actually, as explained in \cite{BrigoBuescu2012}, this formulation of the self-financing condition is an abuse of notation. For example, we have stated that $C_t$ is a unit-of-account of constant value 1, and should therefore obey $dC_t = 0$. However, it should be understood implicitly, when reading \eqref{eq:DifferentialReplicatingPortfolio}, that the differentials of the different components refer to the \textit{gain processes}, including generated dividends. In particular, we will have $dC_t = c_t dt$.}
\begin{equation}
\begin{array}{ll}
dV_t = &\alpha_t dH_t + \beta _t dC_t + \xi_t dCDS^I(t,\, T) + \epsilon_t dCDS^I(t,\,t+dt) + \Omega_t dB(t,\,t+dt)\\
& + \omega_t dB(t,\, T) + \eta_t dCDS^H(t,\,t+dt) \ .
\end{array}
\label{eq:DifferentialReplicatingPortfolio}
\end{equation}

To continue, we expand the differentials of the hedger's short- and long term bonds, including terms corresponding to her jump-to-default, as

\begin{equation}
\left\lbrace
\begin{array}{l}
dB(t,t+dt)=f_{t}^{H}B(t,t+dt)dt + (R_{H}-1)B(t,t+dt)dN_{t}^{H,\mathbb{P}}\\
dB(t,T)=\mathcal{L}_{H}B(t,T)dt+\frac{\partial B(t,T)}{\partial \pi_{t}^{H}} \sigma_{t}^{H}dW_{t}^{H,\mathbb{P}}+\Delta B(t,T)dN_{t}^{H,\mathbb{P}}\\
\end{array}
\right.
\label{eq:BondDynam}
\end{equation}
where $f_{t}^{H}=c_t+\bar{f}_{t}^{H}$ represents the hedger's short term funding rate, and $\bar{f}_{t}^{H}$ is her short term funding spread over the OIS rate $c_t$.

Furthermore, differential changes in short- and long term CDSs written on the investor are as displayed in \cite{Garcia2013}:

\begin{equation}
\left\lbrace
\begin{array}{l}
dCDS^{I}(t,t+dt)=\pi_{t}^{I}dt-(1-R_I)dN_{t}^{I,\mathbb{P}}\\
dCDS^{I}(t,T)=\mathcal{L}_{I}CDS^{I}(t,T)dt+\frac{\partial CDS^{I}(t,T)}{\partial \pi_{t}^{I}} \sigma_{t}^{I}dW_{t}^{I,\mathbb{P}}+\Delta CDS^{I}(t,T)dN_{t}^{I,\mathbb{P}}\\
\end{array}
\right.
\label{eq:CDSIDynam}
\end{equation}
while those of the short term CDS written on the hedger will be given by:

\begin{eqnarray*}
dCDS^{H}(t,t+dt)=\pi_{t}^{H}dt-(1-R_H)dN_{t}^{H,\mathbb{P}}.\\
\end{eqnarray*}

Equating \eqref{eq:dVFromIto} with \eqref{eq:DifferentialReplicatingPortfolio} produces a hedging equation, in which we eliminate the stochastic terms driven by $dW_{t}^{k,\mathbb{P}}$, $k\in\{I,H,S\}$ and $dN_{t}^{I,\mathbb{P}}$, $dN_{t}^{H,\mathbb{P}}$ by taking

\begin{equation}
\begin{array}{cc}
\alpha_{t}=\frac{\frac{\partial V_{t}}{\partial S_{t}}}{\frac{\partial H_{t}}{\partial S_{t}}}& \xi_{t}=\frac{\frac{\partial V_{t}}{\partial \pi_{t}^{I}}}{\frac{\partial CDS^{I}(t,T)}{\partial \pi_{t}^{I}}}\\
\omega_{t}=\frac{\frac{\partial V_{t}}{\partial \pi_{t}^{H}}}{\frac{\partial B(t,T)}{\partial \pi_{t}^{H}}}&\epsilon_{t}=\xi_{t}\frac{\Delta CDS^{I}(t,T)}{1-R_I}-\frac{\Delta V_{t}^{I}}{1-R_I}\\
\multicolumn{2}{c}{\eta_t = -V_t - \frac{\Delta V_{t}^{H}}{1-R_H}}\\
\end{array}
\label{eq: HedgWeights}
\end{equation}

Making use of the PDEs for $H_t$, $CDS^I(t,T)$ and $B(t,T)$, we obtain the final PDE:

\begin{equation}
\hat{\mathcal{L}}_{SIH}V_{t} +\frac{\pi_{t}^{I}}{1-R_I}\Delta V_{t}^{I}+\frac{\pi_{t}^{H}}{1-R_H}\Delta V_{t}^{H}=(f_t^{H}-\pi_{t}^{H})V_{t} \ ,
\label{eq: FinalPDE}
\end{equation}
where

\begin{equation*}
\begin{array}{ll}
\hat{\mathcal{L}}_{SIH}V_{t}=&\frac{\partial V_t}{\partial t}+(r_t-q_t)S_t\frac{\partial V_t}{\partial S_t} + (\mu_t^H-M_t^H\sigma_t^H)\frac{\partial V_t}{\partial \pi_t^H}+ (\mu_t^I-M_t^I\sigma_t^I)\frac{\partial V_t}{\partial \pi_t^I}\\
&+\frac{1}{2}\frac{\partial^2 V_t}{\partial S_t^2}S_t^2(\sigma_t^S)^2)+\frac{1}{2}\frac{\partial^2 V_t}{\partial \pi_t^{H^2}}(\sigma_t^H)^2)+ \frac{1}{2}\frac{\partial^2 V_t}{\partial \pi_t^{I^2}}(\sigma_t^I)^2)\\
&+\frac{\partial^2 V_t}{\partial S_t\partial \pi_t^H}S_t\sigma_t^S\sigma_t^H\rho_t^{S,H}+\frac{\partial^2 V_t}{\partial S_t\partial \pi_t^I}S_t\sigma_t^S\sigma_t^I\rho_t^{S,I}+\frac{\partial^2 V_t}{\partial \pi_t^I\partial \pi_t^H} \sigma_t^I\sigma_t^H\rho_t^{I,H} \ ,
\end{array}
\end{equation*}
$M_t^I$ and $M_t^H$ are the market price of credit risk of the investor and hedger, respectively, that is, the expected excess return of a credit derivative on each of them over the collateral rate divided by the derivatives' volatility.

As shown in Appendix \ref{app:Deriving}, the solution to \eqref{eq: FinalPDE} is given by the following pricing equation:

\begin{equation}
\begin{array}{ll}
V_t=&E_\mathbb{Q}\Big[V_T\exp\Big(-\int_{s=t}^{T}c_s\,ds\Big)|\mathcal{F}_t\Big]\\
&-E_\mathbb{Q}\Big[\int_{s=t}^{T}1_{\{\tau^I>s\}}1_{\{\tau^H>s\}}\exp\Big(-\int_{h=t}^{s}c_h\,dh\Big)\gamma_{s}^{H}V_{s}ds|\mathcal{F}_t\Big]\\
&-E_\mathbb{Q}\Big[\int_{s=t}^{T}\int_{u=s}^{+\infty}\exp\Big(-\int_{v=t}^{s}c_v\,dv\Big)(1-R_I)(V_s^C)^{-}dN_u^{H,\mathbb{Q}}dN_s^{I,\mathbb{Q}}|\mathcal{F}_t\Big]\\
&+E_\mathbb{Q}\Big[\int_{s=t}^{T}\int_{u=s}^{+\infty}\exp\Big(-\int_{v=t}^{s}c_v\,dv\Big)(1-R_H)(V_s^C)^{+}dN_u^{I,\mathbb{Q}}dN_s^{H,\mathbb{Q}}|\mathcal{F}_t\Big]\\
\end{array}
\label{eq: ExpectedValues}
\end{equation}
in a measure $\mathbb{Q}$ in which the drifts of $S_t$, $h_t^H$ and $h_t^I$ are given by $(r_t-q_t)S_t$, $\mu_t^H-M_t^H\sigma_t^H$ and $\mu_t^I-M_t^I\sigma_t^I$, respectively. We have named $\gamma_t^{H}=\bar{f}_t^{H}-\pi_{t}^{H}$, that is, $\gamma_t^{H}$ is the bond-CDS basis. The positive and negative parts of the collateralized value, $(V_t^C)^{+}$ and $(V_t^C)^{-}$, respectively, are defined as absolute values.

The price of the derivative is thus composed of four terms, each expressed in terms of an expected value. The first term is the collateralized value of the derivative, while the last two terms can be identified as CVA and DVA, respectively. It should be noted, from the nested integrations of the jump processes, that both CVA and DVA are based on defaults contingent on the survival of the other party, as follows from our use of a riskless close-out convention. Finally, the second term is an FVA, proportional to the hedger's bond-CDS basis. We have also identified the source of the DVA term as a funding benefit, but will continue to call it DVA due to its interpretation as the CVA that the counterparty would calculate.

Due to the iterative structure of the FVA term, equation \eqref{eq: ExpectedValues} is difficult to solve in general, and as was the case in \cite{PallavaciniBrigo2011}, among others, the solution exhibits no clear separation into a CVA, a DVA and a funding term. In Appendix \ref{app:Depo}, we illustrate these issues by solving the pricing equation for the special case of a cash deposit under idealized conditions. In more general cases a discretization scheme can of course be applied.

\subsection{Price agreement between sophisticated counterparties}

Up to now we have priced the derivative based on the hedger's replication costs. Due to the dependence of the FVA term on the hedger's bond-CDS basis, the price \eqref{eq: ExpectedValues} will be agent-specific, and in general not the same as what would be obtained if the investor were to carry out the replication. We will now briefly consider the case of two sophisticated counterparties, each capable of carrying out the replication, and ask the question of whether they will agree on a price.

Firstly, let us note that the FVA-term is odd in $V_\cdot $, as is the rest of the price equation if we also interchange the CVA and DVA terms, which implies that two sophisticated counterparties will agree on a price if they have the same bond-CDS basis (if $V_t$ is the valuation calculated by the first counterparty, the price obtained by the second counterparty will be obtained by performing $V_t \rightarrow - V_t$, since this will correspond to a solution of his or her pricing equation).

In general, if $V_t^1$ is the valuation obtained by counterparty 1, from her own perspective, and $V_t^2$ is analogously, the valuation as seen by counterparty 2, we see, from the structure of the pricing equation that
\begin{equation}
V_t^1 = F(\gamma^{1}) \ , \\ V_t^2 = -F(\gamma^{2}) \ ,
\end{equation}
for some function $F$ of the bond-CDS basis. Since $F(\gamma^{1})$ is the maximum quantity (possibly negative) which counterparty 1 is willing to pay upfront, upon entering the transaction, while $F(\gamma^{2})$ is the minimum quantity that counterparty 2 is willing to receive, the deal will be closed iff
\begin{equation}
F(\gamma^{2}) \leq F(\gamma^{1}) \ .
\label{eq:DealClosingCondition}
\end{equation}

If we differentiate \eqref{eq: ExpectedValues} with respect to $\gamma^{H}$, holding everything else constant (assuming a deterministic bond-CDS basis), we find
\begin{equation}
\frac{\partial V_t}{\partial \gamma^{H}} = -E_\mathbb{Q}\Big[\int_{s=t}^{T}1_{\{\tau^I>s\}}1_{\{\tau^H>s\}}\exp\Big(-\int_{h=t}^{s}c_h\,dh\Big)V_{s}ds|\mathcal{F}_t\Big] \ .
\end{equation}

In particular, if the expected valuation holds the same sign during the entire life of the derivative, the absolute value of the price will be decreasing in $\gamma$. Together with \eqref{eq:DealClosingCondition} this implies the deal will be closed if and only if the bond-CDS basis of the counterparty which is "creditor" (for which the valuation is positive) is lower or equal to the basis of the "debtor". The same relationship was obtained in \cite{MoriniPrampolini2010} in which the conditions under which a cash deposit could be agreed on were analyzed, taking funding considerations into account. 

\subsection{A steady state expression}

We will now highlight a special result obtained under \eqref{eq: ExpectedValues} (also shown in \cite{MoriniPrampolini2010}) which at first glance might not seem to make economic sense.

Imagine a hedger whose debt is perfectly liquid and has a bond-CDS basis equal to zero. However, due to her CDS she has a funding spread of 200bp above the OIS rate. Let this hedger lend a certain amount of money to a risk-free counterparty\footnote{Like a supranational entity or a multilateral development bank, whose debt is typically weighted by zero when allocating capital for credit risk.}. From the viewpoint of the hedger, there is obviously no DVA in this transaction (there is no change in value upon the hedger's default). There is also no CVA, since the counterparty can be considered as risk-free, and no FVA since the hedger's bond-CDS basis is zero. Therefore, according to \eqref{eq: ExpectedValues}, this transaction is worth the lent amount discounted by the OIS rate, while the funding spread \emph{does not appear} in the valuation formula. Let us give some insight into this.

If the hedger entered this transaction, she would borrow at her financing spread and transfer the money to the risk-free counterparty. Should the hedger default, the counterparty would return the present value of the notional, discounted at the OIS rate (assuming a risk-free closeout), which would then be distributed among the hedger's creditors. Some of these creditors will be new, associated to bonds specifically issued to finance this operation. However, \emph{all} creditors will receive a fraction of this money, raising the recovery of the hedger's debt. This increase in recovery would eventually lead to a funding benefit, reducing her financing costs.

It can be argued that funding rates do not adjust automatically as balance sheets change. In a steady state, however, balance sheets remain constant through time, and funding spreads do reflect the riskiness of companies. This was at the core of \eqref{eq: ExpectedValues}, which relied on the fact that a company would be prepared to pay its own CDS spread for a CDS written on himself, since it reduced its funding spreads by exactly that amount. This reliance on a steady state framework makes \eqref{eq: ExpectedValues} particularly suitable for accountancy purposes. The International Financial Reporting Standard (IFRS) 13 characterizes the fair value of a given instrument as an exit price, that is, the one that would be paid or received in an orderly transaction between market participants. It seems reasonable that this "exit price" is to be considered in a steady state, where prices of particular instruments affect the overall portfolio on a marginal basis.

Apparently, this result seems to be in line with \cite{Hull}, where it is argued that projects that reduce the risk of a company and, therefore, incrementally reduce its funding costs, should be undertaken since they increase shareholder value. However, our analysis introduces two caveats to this conclusion:
\begin{itemize}

\item Liquidity premia, and other effects contained in the bond-CDS basis, cannot be ignored. If the market is charging a company a funding premium beyond what is implied by its riskiness, the value of reducing this riskiness will be diminished. Such considerations were also introduced by the authors of \cite{Hull} in \cite{Hull2012b} and \cite{Hull2013}.

\item Strictly speaking, our pricing equation is valid in a steady state. New deals can thus be priced accordingly if they replace old deals of a similar nature from a credit/funding point of view.

\end{itemize}

\section{Summary and Conclusions}

In this paper we have studied the pricing of an uncollateralized derivatives contract through replication, arguing that, despite not being hedgeable, a definite value can be given to the hedger's jump-to-default component. In fact, it will correspond to supplementing the replicating portfolio with a short-term CDS written on the hedger, who is prepared to pay the market spread for such a product due to the funding benefit it implies. The hedger can carry out a self-financing trading strategy, in which the jump-to-default component will adjust automatically, implying a definite value for the derivatives transaction.

The result is a pricing formula consisting of the value the derivative would have if it were collateralized, together with a bilateral contingent CVA, such as the one first derived in \cite{Brigo2009}, and an FVA term, proportional to the bond-CDS basis. As a consequence, sophisticated counterparties can agree on a price if their bond-CDS bases are related in a certain way. In particular, in deals in which one counterparty acts as a creditor, and the other as a debtor, as in \cite{MoriniPrampolini2010}, the deal will be closed if the creditor has the smaller bond-CDS basis.

It should also be noted that, though not part of our assumptions, the obtained pricing formula is in line with the International Financial Reporting Standard (IFRS) 13 "Fair Value Measurement", which entered into force in January 2013. Largely based on the accounting standard applied in the United States, it intends to harmonize the definition of fair value, characterized as an exit price, that is, the one that would be paid or received in an orderly transaction between market participants. In this setup, \emph{all} cash-flows, including those generated by either party's jump-to-default, must be taken into account. Furthermore, the DVA, equal to the CVA that the counterparty calculates, should clearly be included in an exit price. 

The case of the FVA term is not as clear, being agent-specific since it depends on the hedger's bond-CDS basis. However, if a majority of market participants were to include funding considerations implicitly or explicitly in their pricing, \textit{some} FVA should be included in the calculation of an exit price. In \cite{Hull2013} it is argued that it does not seem to be compatible with IFRS 13 to include entity-specific liquidity considerations into a derivative's fair-value. However, the fair-value of a bond issued by an entity does include such effects, but is still accepted since it is based on market bond prices. Our replicating portfolio consists entirely of market instruments, and is priced consistently with their individual accounting valuations. Since the original deal must be priced in accordance with its replication in order to eliminate arbitrage opportunities, this leads us to conclude that it is correct, from an accounting point of view, to have an entity specific fair-value for a derivatives transaction, based on the understanding that an intrinsic component of the derivative is the entities' issued bonds. In reality, we are simply extending the credit component already present in DVA to encompass the full issued bond price.

Furthermore, in our formulation counterparties with equal bond-CDS bases will arrive at precisely the same price, and the mere fact that a given deal is in the hedger's books would imply the existence of counterparties having the same, or better basis (from the viewpoint of both counterparties since, as shown above, deals closed with different bases will imply a profit to one or both participants). 

An important issue when closing new deals is whether DVA should be (partially) taken into account. In our case, the DVA term has a clear interpretation in terms of a funding benefit, supporting its inclusion. However, funding rates do not adapt instantly, since large portions of issued debt are longer term. The argument presented here would apply to a stable state of the hedger's balance sheet, with a static total DVA generating a stable funding benefit. Still, it could be argued that given the sluggish response in the funding rates to changes in DVA, or, for that matter, to changes in the hedger's risk profile, the hedger may not be willing to offer the full CDS spread $\pi^H_t$ to the investor, in exchange for the implicit protection sold on the hedger's default. This can be seen to be consistent with current market practices in which it is frequent to only recognize partially DVA when closing a deal. For accounting purposes, however, the full DVA should be calculated.

One frequent critique of including DVA in the price charged to counterparties is its negative carry. It would therefore seem to reduce the profitability of the originating desk and/or the CVA desk substantially. However, we would argue that since a stable DVA is directly equivalent to a funding benefit, the institution's treasury, for instance, should compensate DVA originators accordingly\footnote{A way to carry out this compensation is to calculate the notional of the fictitious hedger's CDS present in the replicating portfolio, and book it as an internal transaction between the treasury and the derivatives desk. From the viewpoint of the desk, the replicating portfolio would then be completely comprised of real instruments.}.

In sum, by including a CDS sold by the investor to the hedger in the replicating portfolio we have motivated the inclusion of a DVA term, as well as a residual FVA term, in the pricing equation, which we have argued is compatible with accounting standards. The DVA is identified as a funding benefit, while the residual funding term is governed by the hedger's bond-CDS basis.

\section*{Acknowledgements}


We would like to thank Luis Manuel Garc\'ia Mu\~noz and Juan Antonio de Juan Herrero for helpful comments and suggestions.

\newpage

\appendix

\section{Deriving the Pricing Equation}
\label{app:Deriving}

In this section we will show how the pricing equation \eqref{eq: ExpectedValues} follows from the PDE governing the price process. To do so, we will follow the well-known steps leading to the Feynman-Kac formula (shown in, for example, \cite{KarShreve}). We start with function $V_t=V(t,S_t,\pi_t^I,\pi_t^H)$ that follows the PDE:

\begin{equation*}
\hat{\mathcal{L}}_{SIH}V_{t} +\frac{\pi_{t}^{I}}{1-R_I}\Delta V_{t}^{I}+\frac{\pi_{t}^{H}}{1-R_H}\Delta V_{t}^{H}=(f_t^{H}-\pi_{t}^{H})V_{t} \ ,
\end{equation*}
where
\begin{equation*}
\begin{array}{ll}
\hat{\mathcal{L}}_{SIH}V_{t}=&\frac{\partial V_t}{\partial t}+(r_t-q_t)S_t\frac{\partial V_t}{\partial S_t} + (\mu_t^H-M_t^H\sigma_t^H)\frac{\partial V_t}{\partial \pi_t^H}+ (\mu_t^I-M_t^I\sigma_t^I)\frac{\partial V_t}{\partial \pi_t^I}\\
&+\frac{1}{2}\frac{\partial^2 V_t}{\partial S_t^2}S_t^2(\sigma_t^S)^2)+\frac{1}{2}\frac{\partial^2 V_t}{\partial \pi_t^{H^2}}(\sigma_t^H)^2)+ \frac{1}{2}\frac{\partial^2 V_t}{\partial \pi_t^{I^2}}(\sigma_t^I)^2)\\
&+\frac{\partial^2 V_t}{\partial S_t\partial \pi_t^H}S_t\sigma_t^S\sigma_t^H\rho_t^{S,H}+\frac{\partial^2 V_t}{\partial S_t\partial \pi_t^I}S_t\sigma_t^S\sigma_t^I\rho_t^{S,I}+\frac{\partial^2 V_t}{\partial \pi_t^I\partial \pi_t^H} \sigma_t^I\sigma_t^H\rho_t^{I,H}\\
\end{array}
\end{equation*}
and define the process:
\begin{equation}
X_t=V_t\exp\Big(-\int_{s=0}^{t}c_s\,ds\Big)1_{\{\tau^I>t\}}1_{\{\tau^H>t\}}
\label{eq: ProcessX}
\end{equation}

We then place ourselves in the risk-neutral measure $\mathbb{Q}$ in which the drifts of $S_t$, $\pi_t^H$ and $\pi_t^I$ are given by $(r_t-q_t)S_t$, $\mu_t^H-M_t^H\sigma_t^H$ and $\mu_t^I-M_t^I\sigma_t^I$, respectively. Furthermore, default intensities of the investor and hedger are given by:
\begin{equation*}
\begin{array}{ll}
\lambda_t^{I,\mathbb{Q}}=\frac{\pi_t^I}{1-R_I}& \lambda_t^{H,\mathbb{Q}}=\frac{\pi_t^H}{1-R_H}\\
\end{array}
\end{equation*}
We apply It\^o's Lemma for jump diffusion processes to $X_t$ in $\mathbb{Q}$:
\begin{equation*}
\begin{array}{ll}
dX_t=&\exp\Big(-\int_{s=0}^{t}c_s\,ds\Big)\Big[1_{\{\tau^I>t\}}1_{\{\tau^H>t\}}\Big(-c_tV_tdt+\hat{\mathcal{L}}_{SIH}V_{t} dt+\frac{\partial V_t}{\partial S_t}S_t\sigma_t^SdW_t^S\\
&+\frac{\partial V_t}{\partial \pi_t^I}\sigma_t^IdW_t^I+\frac{\partial V_t}{\partial \pi_t^H}\sigma_t^HdW_t^H\Big)\\
&-1_{\{\tau^I>t\}}V_tdN_t^{H,\mathbb{Q}}-1_{\{\tau^H>t\}}V_tdN_t^{I,\mathbb{Q}}\Big] \ ,\\
\end{array}
\end{equation*}
while from the PDE shown above, we have:

\begin{equation*}
\hat{\mathcal{L}}_{SIH}V_{t}=(f_t^{H}-\pi_{t}^{H})V_{t}-\lambda_t^{I,\mathbb{Q}}\Delta V_{t}^{I}-\lambda_t^{H,\mathbb{Q}}\Delta V_{t}^{H}=(c_t+\bar{f}_t^{H}-\pi_{t}^{H})V_{t}-\lambda_t^{I,\mathbb{Q}}\Delta V_{t}^{I}-\lambda_t^{H,\mathbb{Q}}\Delta V_{t}^{H}
\end{equation*}
so that,

\begin{equation*}
\begin{array}{ll}
dX_t=&\exp\Big(-\int_{s=0}^{t}c_s\,ds\Big)\Big[1_{\{\tau^I>t\}}1_{\{\tau^H>t\}}\Big((\bar{f}_t^{H}-\pi_{t}^{H})V_{t}dt-\lambda_t^{I,\mathbb{Q}}\Delta V_{t}^{I}dt-\lambda_t^{H,\mathbb{Q}}\Delta V_{t}^{H}dt\\
&+\frac{\partial V_t}{\partial S_t}S_t\sigma_t^SdW_t^S+\frac{\partial V_t}{\partial \pi_t^I}\sigma_t^IdW_t^I+\frac{\partial V_t}{\partial \pi_t^H}\sigma_t^HdW_t^H\Big)\\
&+1_{\{\tau^H>t\}}\Delta V_t^IdN_t^{I,\mathbb{Q}}+1_{\{\tau^I>t\}}\Delta V_t^HdN_t^{H,\mathbb{Q}}\\
&-1_{\{\tau^I>t\}}(V_t+\Delta V_t^H)dN_t^{H,\mathbb{Q}}-1_{\{\tau^H>t\}}(V_t+\Delta V_t^I)dN_t^{I,\mathbb{Q}}\Big]\\
\end{array}
\end{equation*}
Naming $\gamma_t^H=\bar{f}_{t}^{H}-\pi_{t}^{H}$, we can integrate between $t$ and $T$, and assuming $\tau^I,\tau^H>t$ we have:

\begin{equation*}
\begin{array}{l}
V_T\exp\Big(-\int_{s=t}^{T}c_s\,ds\Big)1_{\{\tau^I>T\}}1_{\{\tau^H>T\}}-V_t=\\
\int_{s=t}^{T}\exp\Big(-\int_{h=t}^{s}c_h\,dh\Big)\Big[1_{\{\tau^I>s\}}1_{\{\tau^H>s\}}\Big(\gamma_{s}^{H}V_{s}ds\\
-\lambda_s^{I,\mathbb{Q}}\Delta V_{s}^{I}ds-\lambda_s^{H,\mathbb{Q}}\Delta V_{s}^{H}ds+\frac{\partial V_s}{\partial S_s}S_s\sigma_s^SdW_s^S+\frac{\partial V_s}{\partial \pi_s^I}\sigma_s^IdW_s^I+\frac{\partial V_s}{\partial \pi_s^H}\sigma_s^HdW_s^H\Big)\\
+1_{\{\tau^H>s\}}\Delta V_s^IdN_s^{I,\mathbb{Q}}+1_{\{\tau^I>s\}}\Delta V_s^HdN_s^{H,\mathbb{Q}}-1_{\{\tau^I>s\}}(V_s+\Delta V_s^H)dN_s^{H,\mathbb{Q}}\\
-1_{\{\tau^H>s\}}(V_s+\Delta V_s^I)dN_s^{I,\mathbb{Q}}\Big]\\
\end{array}
\end{equation*}
and taking the expectation conditional on $\mathcal{F}_t$

\begin{equation*}
\begin{array}{l}
E_\mathbb{Q}\Big[V_T\exp\Big(-\int_{s=t}^{T}c_s\,ds\Big)1_{\{\tau^I>T\}}1_{\{\tau^H>T\}}|\mathcal{F}_t\Big]-V_t=\\
E_\mathbb{Q}\Big[\int_{s=t}^{T}\exp\Big(-\int_{h=t}^{s}c_h\,dh\Big)\Big(1_{\{\tau^I>s\}}1_{\{\tau^H>s\}}\Big(\gamma_{s}^{H}V_{s}ds\\
-\lambda_s^{I,\mathbb{Q}}\Delta V_{s}^{I}ds-\lambda_s^{H,\mathbb{Q}}\Delta V_{s}^{H}ds+\frac{\partial V_s}{\partial S_s}S_s\sigma_s^SdW_s^S+\frac{\partial V_s}{\partial \pi_s^I}\sigma_s^IdW_s^I+\frac{\partial V_s}{\partial \pi_s^H}\sigma_s^HdW_s^H\Big)\\
+1_{\{\tau^H>s\}}\Delta V_s^IdN_s^{I,\mathbb{Q}}+1_{\{\tau^I>s\}}\Delta V_s^HdN_s^{H,\mathbb{Q}}-1_{\{\tau^I>s\}}(V_s+\Delta V_s^H)dN_s^{H,\mathbb{Q}}\\
-1_{\{\tau^H>s\}}(V_s+\Delta V_s^I)dN_s^{I,\mathbb{Q}}\Big) |\mathcal{F}_t\Big]\\
\end{array}
\end{equation*}

On the right hand side, the expected values of the terms in $dW_s^k$, $k\in\{S, I, H\}$ are zero since they represent the expected values of It\^o integrals. Furthermore, for any function $f(s)$ the definition of default intensity implies that\footnote{The default intensity $\lambda_t$ is equal to $\frac{\phi(t)}{P(\tau > t)}$, where $\phi(t)$ is the default density. Integrating over $\tau$ in the expected value allows us to substitute $1_{\{\tau>s\}}$ for $P(\tau > t)$, and both sides of equation \eqref{eq:DefaultIntensityIdentity} will therefore correspond to integrals of $f(t)$ multiplied by the default density.}:

\begin{equation}
E_\mathbb{Q}\Big[\int_{s=t}^{T}1_{\{\tau>s\}}\lambda_sf(s)ds\Big]=E_\mathbb{Q}\Big[\int_{s=t}^{T}f(s)dN_{s}\Big] \ ,
\label{eq:DefaultIntensityIdentity}
\end{equation}
allowing us to write

\begin{equation}
\begin{array}{ll}
V_t=& E_\mathbb{Q}\Big[V_T\exp\Big(-\int_{s=t}^{T}c_s\,ds\Big)1_{\{\tau^I>T\}}1_{\{\tau^H>T\}}|\mathcal{F}_t\Big]\\
&-E_\mathbb{Q}\Big[\int_{s=t}^{T}1_{\{\tau^I>s\}}1_{\{\tau^H>s\}}\exp\Big(-\int_{h=t}^{s}c_h\,dh\Big)\gamma_s^{H}V_{s}ds|\mathcal{F}_t\Big]\\
&+E_\mathbb{Q}\Big[\int_{s=t}^{T}1_{\{\tau^H>s\}}\exp\Big(-\int_{h=t}^{s}c_h\,dh\Big)(V_s+\Delta V_{s}^{I})dN_s^{I,\mathbb{Q}}|\mathcal{F}_t\Big]\\
&+E_\mathbb{Q}\Big[\int_{s=t}^{T}1_{\{\tau^I>s\}}\exp\Big(-\int_{h=t}^{s}c_h\,dh\Big)(V_s+\Delta V_{s}^{H})dN_s^{H,\mathbb{Q}}|\mathcal{F}_t\Big]\\
\end{array}
\label{eq:FirstExpect}
\end{equation}

Now, as in \cite{Brigo2009}, we can define the following (mutually exclusive and exhaustive)
events ordering the default times:

\begin{eqnarray*}
A = \{\tau_{I}\leq\tau_{H}\leq T\}&&E = \{T \leq\tau_{I}\leq \tau_{H}\}\\
B = \{\tau_{I}\leq T \leq \tau_{H}\}&&F = \{T \leq\tau_{H}\leq \tau_{I}\}\\
C = \{\tau_{H}\leq\tau_{I}\leq T\}&&\\
D = \{\tau_{H}\leq T \leq \tau_{I}\}&&\\
\end{eqnarray*}
Notice that A through D are the default events, while E and F are the
non-default ones. Using this notation we can rewrite

\begin{equation}
\begin{array}{l}
E_\mathbb{Q}\Big[V_T\exp\Big(-\int_{s=t}^{T}c_s\,ds\Big)1_{\{\tau^I>T\}}1_{\{\tau^H>T\}}|\mathcal{F}_t\Big]=E_\mathbb{Q}\Big[V_T\exp\Big(-\int_{s=t}^{T}c_s\,ds\Big)1_{\{E\cup F\}}|\mathcal{F}_t\Big]=\\
E_\mathbb{Q}\Big[V_T\exp\Big(-\int_{s=t}^{T}c_s\,ds\Big)|\mathcal{F}_t\Big]-E_\mathbb{Q}\Big[V_T\exp\Big(-\int_{s=t}^{T}c_s\,ds\Big)1_{\{A\cup B\}}|\mathcal{F}_t\Big]\\
-E_\mathbb{Q}\Big[V_T\exp\Big(-\int_{s=t}^{T}c_s\,ds\Big)1_{\{C\cup D\}}|\mathcal{F}_t\Big]\\
\end{array}
\end{equation}

Furthermore,

\begin{equation}
\begin{array}{l}
E_\mathbb{Q}\Big[V_T\exp\Big(-\int_{s=t}^{T}c_s\,ds\Big)1_{\{A\cup B\}}|\mathcal{F}_t\Big]\\
=E_\mathbb{Q}\Big[\int_{s=t}^{T}\int_{u=s}^{+\infty}V_T\exp\Big(-\int_{v=t}^{T}c_v\,dv\Big)dN_u^{H,\mathbb{Q}}dN_s^{I,\mathbb{Q}}|\mathcal{F}_t\Big]\\
=E_\mathbb{Q}\Big[E_\mathbb{Q}\Big[\int_{s=t}^{T}\int_{u=s}^{+\infty}V_T\exp\Big(-\int_{v=t}^{T}c_v\,dv\Big)dN_u^{H,\mathbb{Q}}dN_s^{I,\mathbb{Q}}|\mathcal{F}_s\Big]|\mathcal{F}_t\Big]\\
=E_\mathbb{Q}\Big[\int_{s=t}^{T}dN_s^{I,\mathbb{Q}}E_\mathbb{Q}\Big[\int_{u=s}^{+\infty}V_T\exp\Big(-\int_{v=t}^{T}c_v\,dv\Big)dN_u^{H,\mathbb{Q}}|\mathcal{F}_s\Big]|\mathcal{F}_t\Big]\\
=E_\mathbb{Q}\Big[\int_{s=t}^{T}dN_s^{I,\mathbb{Q}}E_\mathbb{Q}\Big[V_T\exp\Big(-\int_{v=t}^{T}c_v\,dv\Big)1_{\{\tau^H>s\}}|\mathcal{F}_s\Big]|\mathcal{F}_t\Big]\\
=E_\mathbb{Q}\Big[\int_{s=t}^{T}1_{\{\tau^H>s\}}dN_s^{I,\mathbb{Q}}E_\mathbb{Q}\Big[V_T\exp\Big(-\int_{v=t}^{T}c_v\,dv\Big)|\mathcal{F}_s\Big]|\mathcal{F}_t\Big]\\
=E_\mathbb{Q}\Big[\int_{s=t}^{T}1_{\{\tau^H>s\}}dN_s^{I,\mathbb{Q}}V_s^C\exp\Big(-\int_{v=t}^{s}c_v\,dv\Big)|\mathcal{F}_t\Big]\\
\end{array}
\end{equation}
where $V_t^C$ represents the time $t$ value that the derivative would have if it were completely collateralized.

In a similar fashion, we can obtain

\begin{equation}
\begin{array}{l}
E_\mathbb{Q}\Big[V_T\exp\Big(-\int_{s=t}^{T}c_s\,ds\Big)1_{\{C\cup D\}}|\mathcal{F}_t\Big]\\
=E_\mathbb{Q}\Big[\int_{s=t}^{T}1_{\{\tau^I>s\}}dN_s^{H,\mathbb{Q}}V_s^C\exp\Big(-\int_{v=t}^{s}c_v\,dv\Big)|\mathcal{F}_t\Big]\\
\end{array}
\end{equation}
Returning to (\ref{eq:FirstExpect}), we substitute terms and get:

\begin{equation}
\begin{array}{ll}
V_t=&E_\mathbb{Q}\Big[V_T\exp\Big(-\int_{s=t}^{T}c_s\,ds\Big)|\mathcal{F}_t\Big]\\
&-E_\mathbb{Q}\Big[\int_{s=t}^{T}1_{\{\tau^I>s\}}1_{\{\tau^H>s\}}\exp\Big(-\int_{h=t}^{s}c_h\,dh\Big)\gamma_s^{H}V_{s}ds|\mathcal{F}_t\Big]\\
&+E_\mathbb{Q}\Big[\int_{s=t}^{T}1_{\{\tau^H>s\}}\exp\Big(-\int_{v=t}^{s}c_v\,dv\Big)(V_s+\Delta V_s^I-V_s^C)dN_s^{I,\mathbb{Q}}|\mathcal{F}_t\Big]\\
&+E_\mathbb{Q}\Big[\int_{s=t}^{T}1_{\{\tau^I>s\}}\exp\Big(-\int_{v=t}^{s}c_v\,dv\Big)(V_s+\Delta V_s^H-V_s^C)dN_s^{H,\mathbb{Q}}|\mathcal{F}_t\Big]\\
\end{array}
\end{equation}
Now, if we assume, in accordance with a riskless close-out, that after the investor's default, $V_s$ jumps to $R_IV_s^C$ if $V_s^C<0$ and to $V_s^C$ if $V_s^C\geq 0$, and that after the hedger's default $V_s$ jumps to $V_s^C$ if $V_s^C<0$ and to $R_HV_s^C$ if $V_s^C\geq 0$, then

\begin{equation}
\begin{array}{ll}
V_t=&E_\mathbb{Q}\Big[V_T\exp\Big(-\int_{s=t}^{T}c_s\,ds\Big)|\mathcal{F}_t\Big]\\
&-E_\mathbb{Q}\Big[\int_{s=t}^{T}1_{\{\tau^I>s\}}1_{\{\tau^H>s\}}\exp\Big(-\int_{h=t}^{s}c_h\,dh\Big)\gamma_s^{H}V_{s}ds|\mathcal{F}_t\Big]\\
&+E_\mathbb{Q}\Big[\int_{s=t}^{T}\int_{u=s}^{+\infty}\exp\Big(-\int_{v=t}^{s}c_v\,dv\Big)(1-R_I)(V_s^C)^{-}dN_u^{H,\mathbb{Q}}dN_s^{I,\mathbb{Q}}|\mathcal{F}_t\Big]\\
&-E_\mathbb{Q}\Big[\int_{s=t}^{T}\int_{u=s}^{+\infty}\exp\Big(-\int_{v=t}^{s}c_v\,dv\Big)(1-R_H)(V_s^C)^{+}dN_u^{I,\mathbb{Q}}dN_s^{H,\mathbb{Q}}|\mathcal{F}_t\Big]\\
\end{array}
\end{equation}

\newpage

\section{Pricing a Cash Deposit}

\label{app:Depo}

In this section we will present the solution to the pricing equation \eqref{eq: ExpectedValues} for the simple case of a Cash Deposit, assuming constant default hazard rates $\lambda^I$ and $\lambda^H$ for the investor (lender) and the hedger (borrower), respectively, as well as constant bond-CDS bases and a constant OIS rate. At the end we will also state the solution in the case of time-varying parameters.

We recall the pricing equation:

\begin{equation*}
\begin{array}{ll}
V_t=&E_\mathbb{Q}\Big[V_T\exp\Big(-\int_{s=t}^{T}c_s\,ds\Big)|\mathcal{F}_t\Big]\\
&-E_\mathbb{Q}\Big[\int_{s=t}^{T}1_{\{\tau^I>s\}}1_{\{\tau^H>s\}}\exp\Big(-\int_{h=t}^{s}c_h\,dh\Big)\gamma_s^{H}V_{s}ds|\mathcal{F}_t\Big]\\
&+E_\mathbb{Q}\Big[\int_{s=t}^{T}\int_{u=s}^{+\infty}\exp\Big(-\int_{v=t}^{s}c_v\,dv\Big)(1-R_I)(V_s^C)^{-}dN_u^{H,\mathbb{Q}}dN_s^{I,\mathbb{Q}}|\mathcal{F}_t\Big]\\
&-E_\mathbb{Q}\Big[\int_{s=t}^{T}\int_{u=s}^{+\infty}\exp\Big(-\int_{v=t}^{s}c_v\,dv\Big)(1-R_H)(V_s^C)^{+}dN_u^{I,\mathbb{Q}}dN_s^{H,\mathbb{Q}}|\mathcal{F}_t\Big]\ . \\
\end{array}
\end{equation*}
For simplicity, let us assume that interest rates, CDS spreads, recovery, default intensities and funding spreads are constant. These assumptions leave us the following expression:

\begin{equation}
\begin{array}{ll}
V_t=&e^{-c(T-t)}E_\mathbb{Q}[V_T]-E_\mathbb{Q}\Big[\int_{s=t}^{T}e^{-\lambda_I(s-t)}e^{-\lambda_H(s-t)}e^{-c(s-t)}\gamma^{H}V_{s}ds\Big]\\
&+(1-R_I)E_\mathbb{Q}\Big[\int_{s=t}^{T}\Big(\lambda_Ie^{-\lambda_I(s-t)}\Big)e^{-\lambda_H(s-t)}e^{-c(s-t)}(V_s^C)^{-}ds\Big]\\
&-(1-R_H)E_\mathbb{Q}\Big[\int_{s=t}^{T}e^{-\lambda_I(s-t)}\Big(\lambda_He^{-\lambda_H(s-t)}\Big)e^{-c(s-t)}(V_s^C)^{+}ds\Big]\ . \\
\end{array}
\end{equation}

We are going to compute the solution for a Cash Deposit of notional N with a single cash-flow at maturity T where the investor will act as the lender and the hedger as the borrower. Therefore,

\begin{itemize}
\item $V_T=N$
\item $(V_s^C)^{+}=Ne^{-c(T-s)}$
\item $(V_s^C)^{-}=0$
\end{itemize}
The equation that must be solved is therefore:

\begin{equation}
\begin{array}{ll}
V_t=&e^{-c(T-t)}N-\gamma^{H}\int_{s=t}^{T}e^{-(\lambda_I+\lambda_H+c)(s-t)}V_{s}ds\\
&-(1-R_H)N\lambda_He^{-c(T-t)}\int_{s=t}^{T}e^{-(\lambda_I+\lambda_H)(s-t)}ds \ . \\
\end{array}
\end{equation}
If we evaluate the latter integral and multiply the whole equation by $\frac{e^{c(T-t)}}{N}$, we get

\begin{equation}
\begin{array}{ll}
e^{c(T-t)}\frac{V_{t}}{N}=&1-\gamma^{H}\int_{s=t}^{T}e^{-(\lambda_I+\lambda_H)(s-t)}e^{c(T-s)}\frac{V_{s}}{N}ds\\
&-(1-R_H)\frac{\lambda_H}{\lambda_I+\lambda_H}\Big[1-e^{-(\lambda_I+\lambda_H)(T-t)}\Big]\ . \\
\end{array}
\end{equation}

Let us define $V_{s}^{*}=e^{c(T-s)}\frac{V_s}{N}$. Then we have

\begin{equation}
\begin{array}{ll}
V_{t}^{*}=&1-\gamma^{H}\int_{s=t}^{T}e^{-(\lambda_I+\lambda_H)(s-t)}V_{s}^{*}ds\\
&-(1-R_H)\frac{\lambda_H}{\lambda_I+\lambda_H}\Big[1-e^{-(\lambda_I+\lambda_H)(T-t)}\Big]\ . \\
\end{array}
\end{equation}

Performing a change of variables by defining $\tilde{t}=(\lambda_I+\lambda_H)(T-t)$ and $\tilde{s}=(\lambda_I+\lambda_H)(T-s)$, together with the notation $\tilde{V}_{\tilde{t}}=V_t^{*}$, we get

\begin{equation}
\begin{array}{l}
\tilde{V}_{\tilde{t}}=1-\frac{\gamma^{H}}{\lambda_I+\lambda_H}e^{-\tilde{t}}\int_{\tilde{s}=0}^{\tilde{t}}e^{\tilde{s}}\tilde{V}_{\tilde{s}}d\tilde{s}-(1-R_H)\frac{\lambda_H}{\lambda_I+\lambda_H}\Big(1-e^{-\tilde{t}}\Big)\ . \\
\end{array}
\end{equation}
If we now multiply all terms by $e^{\tilde{t}}$ and differentiate with respect to $\tilde{t}$, we have transformed the integral equation into the ordinary differential equation

\begin{equation}
\begin{array}{l}
\tilde{V}_{\tilde{t}}^{'}+\Big(1+\frac{\gamma^{H}}{\lambda_I+\lambda_H}\Big)\tilde{V}_{\tilde{t}}=1-(1-R_H)\frac{\lambda_H}{\lambda_I+\lambda_H}\ . \\
\end{array}
\end{equation}

Solving this equation and undoing previous changes of variables we find

\begin{equation}
\begin{array}{l}
e^{c(T-t)}\frac{V_t}{N}=\Big[1-(1-R_H)\frac{\lambda_H}{\lambda_I+\lambda_H}\Big]\Big[\frac{\lambda_I+\lambda_H}{\lambda_I+\lambda_H+\gamma^H}\Big]+\bar{K}e^{-(\lambda_I+\lambda_H+\gamma^H)(T-t)} , \\
\end{array}
\label{eq:SolvDifEq}
\end{equation}
where $\bar{K}$ is an integration constant. Imposing the terminal condition $V_T=N$, it needs to be

\begin{equation}
\begin{array}{l}
\bar{K}=1-\Big[1-(1-R_H)\frac{\lambda_H}{\lambda_I+\lambda_H}\Big]\Big[\frac{\lambda_I+\lambda_H}{\lambda_I+\lambda_H+\gamma^H}\Big] \ . \\
\end{array}
\end{equation}

Substituting back in (\ref{eq:SolvDifEq}) and rearranging terms, we get the final solution.

\begin{equation}
\begin{array}{ll}
V_t&=Ne^{-c(T-t)}\Big[1-\frac{(1-R_H)\lambda_H+\gamma^H}{\lambda_I+\lambda_H+\gamma^H}\Big(1-e^{-(\lambda_I+\lambda_H+\gamma^H)(T-t)}\Big)\Big] \ . \\ 
\end{array}
\label{eq:DepositFinalPrice}
\end{equation}

It should be noted that in the case of no credit risk ($\lambda_H = \lambda_I = 0$) the valuation becomes
\begin{equation}
V_t = Ne^{-\left(c +\gamma^H \right) (T-t)} \ ,
\end{equation}
with the interpretation that the discount rate gets an additional liquidity contribution given by the bond-CDS basis, in accordance with the results of \cite{MoriniPrampolini2010}. However, if we set $R_H=0$ in \eqref{eq:DepositFinalPrice} we are left with
\begin{equation}
\begin{array}{ll}
V_t&=Ne^{-c(T-t)}\Big[1-\frac{\lambda_H+\gamma^H}{\lambda_I+\lambda_H+\gamma^H}\Big(1-e^{-(\lambda_I+\lambda_H+\gamma^H)(T-t)}\Big)\Big] \\
&=Ne^{-c(T-t)}\Big[\frac{\lambda_I}{\lambda_I+\lambda_H+\gamma^H}+ \Big(1- \frac{\lambda_I}{\lambda_I+\lambda_H+\gamma^H} \Big) e^{-(\lambda_I+\lambda_H+\gamma^H)(T-t)}\Big] \ ,
\end{array}
\label{eq:DepositFinalPriceZeroRec}
\end{equation}
to be compared with the equation
\begin{equation}
V_t = Ne^{-\left(c + \lambda_H + \gamma^H \right) (T-t)} \ 
\end{equation}
that is obtained from \cite{MoriniPrampolini2010} (equation (11)) in the zero-recovery case (translated to our notation). We see that the two equations only agree if the investor is default-free ($\lambda_I = 0$). The reason for the discrepancy in general is that \eqref{eq:DepositFinalPriceZeroRec} takes into account that the default of the lender will require the borrower to pre-emptively return the notional, discounted at the riskless rate, while \cite{MoriniPrampolini2010} does not include this effect. The value to the borrower will therefore depend on the credit-quality of the lender. Indeed, let us take the limit $\lambda_I \rightarrow \infty$ in \eqref{eq:DepositFinalPriceZeroRec}, implying an imminent default of the investor, while keeping the hedger's credit state constant. We then obtain precisely the close-out amount.

Let us end by stating (the interested reader can attempt its derivation as an exercise) a generalization of \eqref{eq:DepositFinalPrice} to the case where the pricing parameters can vary in time (albeit deterministically), since the result has a rather interesting structure. Letting $\lambda_H(t)$, $\lambda_I(t)$ and $\gamma^H(t)$ denote the time-varying parameters, and $D(t,\, t')$ the riskless discount curve, the result is
\begin{equation}
\begin{array}{ll}
V_t=N D(t,\,T)&\Big[1-\int_t^T\big((1-R_H) \lambda_H(s)+\gamma^H(s)\big)\times \\
& \hspace{50pt} \exp \left(-\int_t^s(\lambda_I(u)+\lambda_H(u)+\gamma^H(u))\, du \right) \, ds\Big] \ . \\ 
\end{array}
\label{eq:DepositFinalPriceTimeVarying}
\end{equation}
The equation implies a credit and funding correction corresponding to a given future time $s$ as basically the hedger's bond short-rate, as a spread over the OIS rate, $(1-R_H)\lambda_H(s)+\gamma^H(s)$, weighted by a liquidity-adjusted survival probability.

\end{document}